\documentclass[twocolumn,showpacs,preprintnumbers,amsmath,amssymb]{revtex4}
\usepackage{graphicx}
\usepackage{dcolumn}
\usepackage{bm}
\newcommand{\bef}{\begin{figure}}
\newcommand{\eef}{\end{figure}}

\newcommand{\be}{\begin{equation}}
\newcommand{\ee}{\end{equation}}
\newcommand{\bea}{\begin{eqnarray}}
\newcommand{\eea}{\end{eqnarray}}

\begin{document}

\title{Systematic study of the elliptic flow parameter using a heavy-ion collision model}



\author{Md. Nasim and
        Bedangadas Mohanty 
}

\affiliation{School of Physical Sciences, National Institute of Science Education and Research, Bhubaneswar-751005, India }

\begin{abstract}
Elliptic flow parameter, $v_{2}$ is consider as a sensitive probe for early dynamics of
the heavy-ion collision. In this work we have discussed the effect of detector
efficiency, procedure of centrality determination, effect of resonance
decay, procedure to obtain event plane resolution on
the measured $v_{2}$ by standard event plane method within the framework of a transport
model. The measured value of $v_{2}$ depends on the detector
efficiency in particle number counting. The effect of centrality
determination is found to be negligible. The method of event-by-event
correction of event plane resolution for wide centrality bin yields in
results closer to the true value of $v_{2}$. The effect of resonance decay
is seen to decrease the $v_{2}$ of $\pi$, $K$ and $p$. We also propose
a procedure to correct for an event bias effect on $v_{2}$ while
comparing the minimum bias centrality $v_{2}$  values for different
multi-strange hadrons. Finally we have
presented a model based confirmation of the recently proposed relation
between $v_{2}$ obtained using event plane method and scalar product
method to the true value of $v_{2}$.
\end{abstract}
\maketitle

\section{Introduction}
\label{intro}
The study of the azimuthal angle distribution for hadrons produced in
high energy heavy-ion collisions is considered as a very useful observable for
understanding the properties of the hot and dense matter formed in the
collisions~\cite{ATLAS:2011ah,Chatrchyan:2012xq,Aamodt:2010pa,ALICE:2011ab,Ackermann:2000tr,Adler:2001nb,Adler:2003kt,Adare:2006nq}. 
The second Fourier coefficient of the azimuthal angle
($\phi$) distribution of the particles produced in heavy-ion
collision with respect to the reaction plane angle ($\Psi$) is called
as the elliptic flow parameter ($v_{2}$)~\cite{art}. It is defined as 
\begin{equation}
v_{2}=\langle\langle\cos(2(\phi-\Psi))\rangle \rangle=\langle\langle\frac{p_x^2-p_y^2}{p_x^2+p_y^2}\rangle\rangle,
\end{equation}
where $p_x$ and $p_y$ are the $x$ and $y$ components of the particle
momenta, respectively. The $\langle \langle$  $\rangle \rangle$ denotes the average
over all particles in all events. The $\Psi$ is an estimate of the angle subtended by
the plane formed by impact parameter and beam ($z$) axis with respect to the $x$-axis.
The magnitude of $v_{2}$ is found to be sensitive to the equation of
state, thermalization, transport coefficients of the medium, and initial conditions
in the heavy-ion collisions~\cite{flow1,flow2,Romatschke:2007mq}.

Since $v_{2}$ is an important observable in high energy heavy-ion
collisions, it is necessary to study the effect of various experimental
conditions towards the measurement of its true value. In this work we
study the effect of finite particle number counting (detector
efficiency), method of centrality determination, effect of correcting
for finite resolution in obtaining $\Psi$ and effect of resonances. For this
study we use transport based models, which provides an ideal set up to
study and quantify the above effects. From the models we know the true
value of $v_{2}$ at the same time it provides information of particles
produced in heavy-ion collisions that can be treated in a manner
similar to the actual experimental conditions. For minimum bias collisions
the event class (in terms of average initial spatial anisotropy)
having a rare heavy particle like $\Omega$ baryon could be
different from that having a proton (copious produced). This may
necessitate an appropriate correction to the measured $v_{2}$ of different hadrons for wide
centrality class (0-80\%) so that they can be compared among themselves. 
In addition, we also use
this model framework to verify the recently proposed relation
that $v_{2}$ obtained using event plane method approaches 
the root-mean square $v_{2}$  (as obtained from the scalar product
method) in the small event plane resolution limit while it approaches
the mean $v_{2}$ value in the high event plane resolution limit~\cite{Luzum:2012da}.

The paper is organised as follows.
In section~\ref{sec:1}, we describe the transport based models used in this study.
The effect of detector efficiency on measured $v_{2}$ is presented
in section~\ref{sec:2}. In section~\ref{sec:3}, we investigate the effect of centrality
selection procedure on measured $v_{2}$. The event plane resolution 
correction methods, event bias correction method and effect of resonance decay on $v_{2}$ has been
discussed in sections~\ref{sec:4},~\ref{sec:5} and~\ref{sec:6} respectively. In section~\ref{sec:7} we discuss
the recently proposed relation between $v_{2}$ obtained from event
plane and scalar product method. Finally section~\ref{sec:8} summarises our
findings. 

\section{Model Description}
\label{sec:1}
The A Multi Phase Transport (AMPT) model~\cite{ampt} (version: 25t7d) uses the same
initial conditions as in HIJING~\cite{hijing}. However the minijet
partons are made to undergo scattering before they are allowed
to fragment into hadrons. The string melting (SM) version of the AMPT model 
is based on the idea that for energy densities beyond 
a critical value of $\sim$ 1 GeV/$\rm {fm}^{3}$, it is difficult to
visualize the coexistence of strings (or hadrons) and partons.  Hence
the need to melt the strings to partons. This is done by converting the mesons to
a quark and anti-quark pair, baryons to three quarks etc.  The scattering of the quarks
are based on parton cascade~\cite{ampt}. Once the interactions stop, the partons then 
hadronizes through the mechanism of parton coalescence. The interactions between
the minijet partons in AMPT model and those between partons in the AMPT-SM
model could give rise to substantial $v_{2}$. All results presented
here using this model is for Au+Au collisions at $\sqrt{s_{NN}}$ = 200
GeV at midrapidity (-0.5 $<$ $\eta$ $<$ 0.5) with total event statistics of
1.8 million minimum bias (0-80\%) events. The true event plane angle is fixed to zero degree for
the simulation results presented. 

The Ultra relativistic Quantum 
Molecular Dynamics (UrQMD) model (version: 3.3) is based on a microscopic transport theory where the
phase space description of the reactions are important~\cite{urqmd}. It allows for the propagation
of all hadrons on classical trajectories in combination with stochastic binary 
scattering, color string formation and resonance decay. It incorporates baryon-baryon,
meson-baryon and meson-meson interactions, the collisional term includes more than 50 
baryon species and 45 meson species. This model is used to understand
the resonance decay effect on measured $v_{2}$. The analysis makes use
of 1.5 million minimum bias (0-80\%) events for Au+Au collisions at
$\sqrt{s_{NN}}$ = 200 GeV at midrapidity.

\section{Effect of detector efficiency}
\label{sec:2}
\bef
\begin{center}
\includegraphics[scale=0.40]{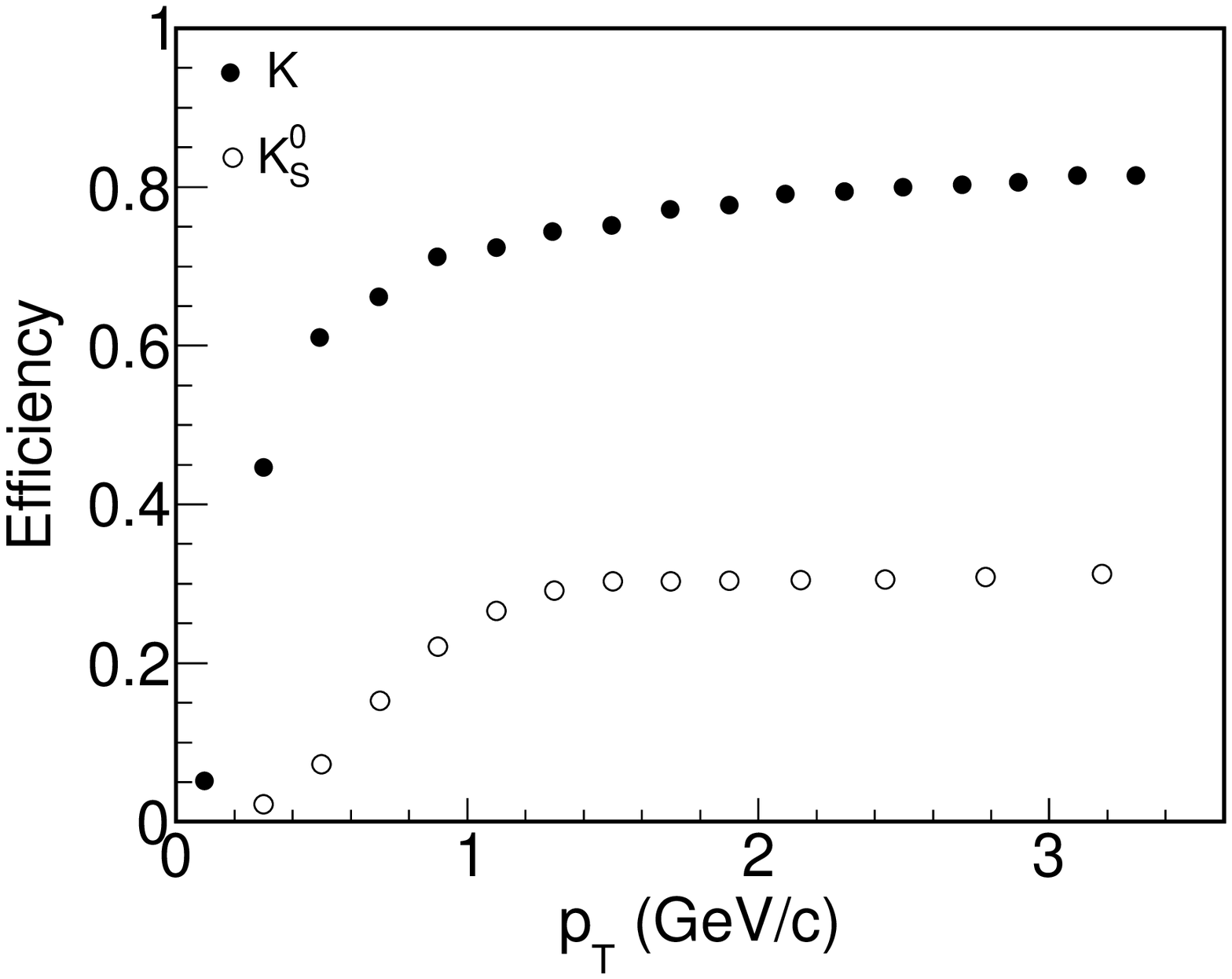}
\caption{Typical particle reconstruction efficiency as function
of $p_{T}$ for charged kaon and $K^{0}_{S}$ at midrapidity in Au+Au collisions at
 high energy. }
\label{fig1}
\end{center}
\eef
\bef
\begin{center}
\includegraphics[scale=0.43]{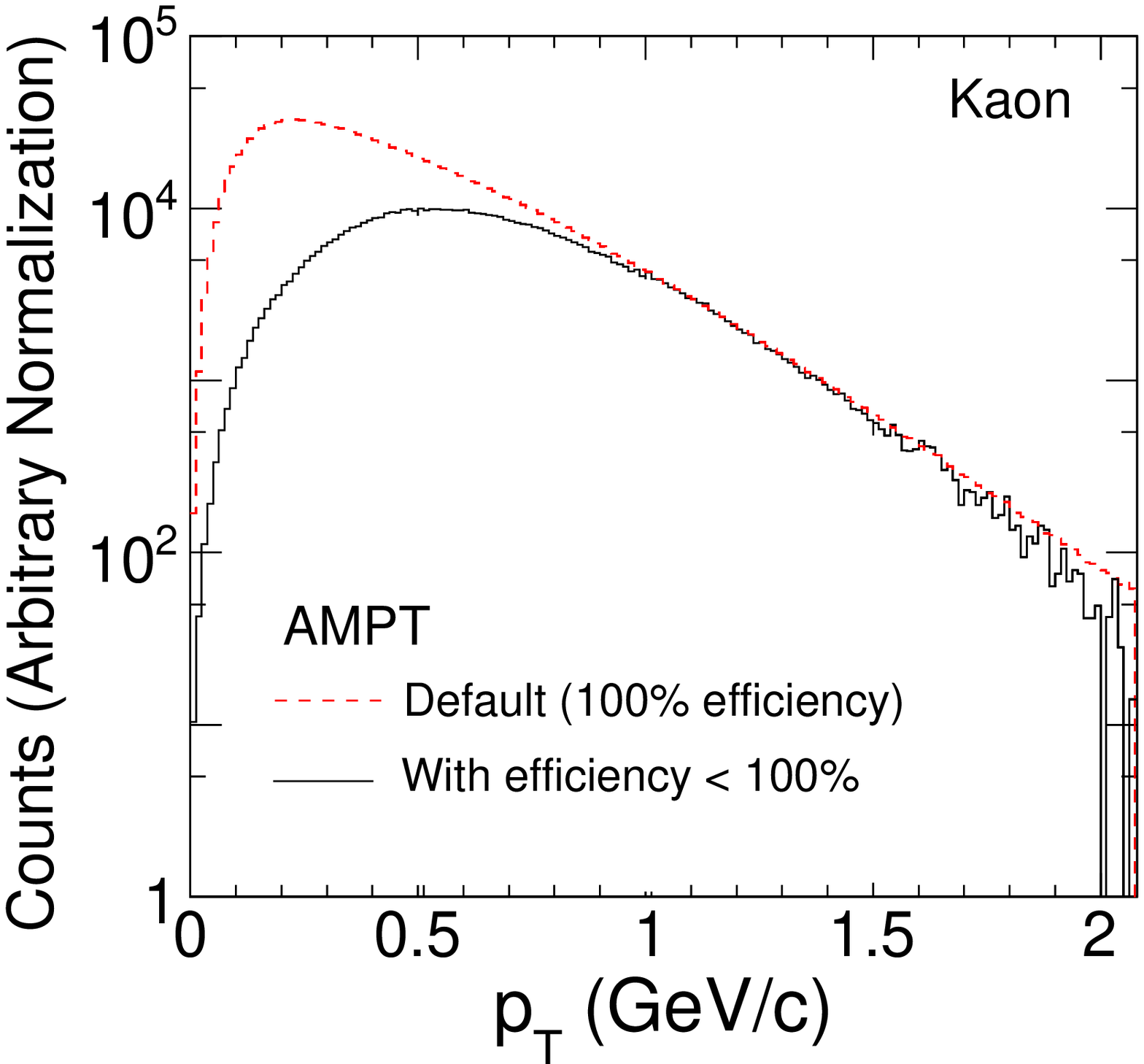}
\caption{(Color online) Kaon yield  as function of $p_{T}$ from AMPT
  model for Au+Au collisions at $\sqrt{s_{NN}}$ = 200 GeV. Red
  dash line corresponds to the yield of kaon obtained directly from the
  AMPT model and black solid line corresponds to the kaon yield after
  incorporating the finite $K^{0}_{S}$ reconstruction efficiency
  values as shown in Fig.~\ref{fig1}. Yields of the two spectra are
  normalised at $p_{T}$ = 1 GeV/$c$.}
\label{fig2}
\end{center}
\eef
In this section we discuss the effect of detector efficiency
on measured $v_{2}$ in a typical heavy-ion experiment.  The AMPT
model is used to simulate Au+Au collision and introduced the effect of finite detector
efficiency on particle number counting.  A realistic detector
efficiency for reconstruction of charged kaon and $K^{0}_{S}$ as a
function of the transverse momentum ($p_{T}$) of the
measured hadron from a  typical heavy-ion experiment~\cite{Aggarwal:2010ig,Abelev:2008ab} as shown in
Fig.~\ref{fig1} is considered.  The reconstruction efficiency for neutral kaons,
which have to be reconstructed from the decayed charged pion daughters
are  smaller compared to those from more directly
reconstructed charged kaons. Two sets of events were considered, one
with 100\% reconstruction efficiency versus $p_{T}$ termed as
``default'' and the other where the kaon tracks in a given event for a
given $p_{T}$ range were randomly removed as per the efficiency shown
in Fig.~\ref{fig1}. The resultant yield of the charged kaons from AMPT model
with $K^{0}_{S}$ reconstruction efficiency effect incorporated has been compared to the default
case in Fig.~\ref{fig2}. The two distributions have been normalised at
their respective yield values at $p_{T}$ = 1 GeV/$c$. The effect of
finite particle reconstruction efficiency can be seen from the shape of yield
vs. $p_{T}$ distribution below $p_{T}$ = 1 GeV/$c$ and as the efficiency
values are constant with $p_{T}$ beyond 1 GeV/$c$ the spectra shape
are similar at high $p_{T}$ (consistent with Fig.~\ref{fig1}).

The elliptic flow of charged kaon has been calculated for
three different condition: (a) with 100$\%$ particle reconstruction
efficiency (labeled as default),  (b) with charged kaon reconstruction
efficiency and (c) with $K^{0}_{S}$ reconstruction
efficiency. Figure~\ref{fig3}  shows the comparison of kaon $v_{2}$
for the above three different cases. The kaon $v_{2}$ from AMPT for
default case, with charged kaon
reconstruction efficiency and with $K^{0}_{S}$ reconstruction
efficiency are shown as solid black circle, open blue square and open
red circle, respectively. The bottom panel of Fig.~\ref{fig3}  shows
the ratios of default kaon $v_{2}$ to that obtained with two different particle reconstruction
efficiency.  There is a change due to the
finite particle reconstruction efficiency on the measured $v_{2}$. The
change in $v_{2}$ due to $K^{0}_{S}$ reconstruction efficiency is
about 10$\%$ to 30$\%$ while for the results using the higher charged kaon
reconstruction efficiency the change is less than 5$\%$. Our study
shows a need for correcting the effect of difference in finite reconstruction
efficiency between charged kaon and $K^{0}_{S}$, before the measured
$v_{2}$ values are compared in the experiments.

\bef
\begin{center}
\includegraphics[scale=0.40]{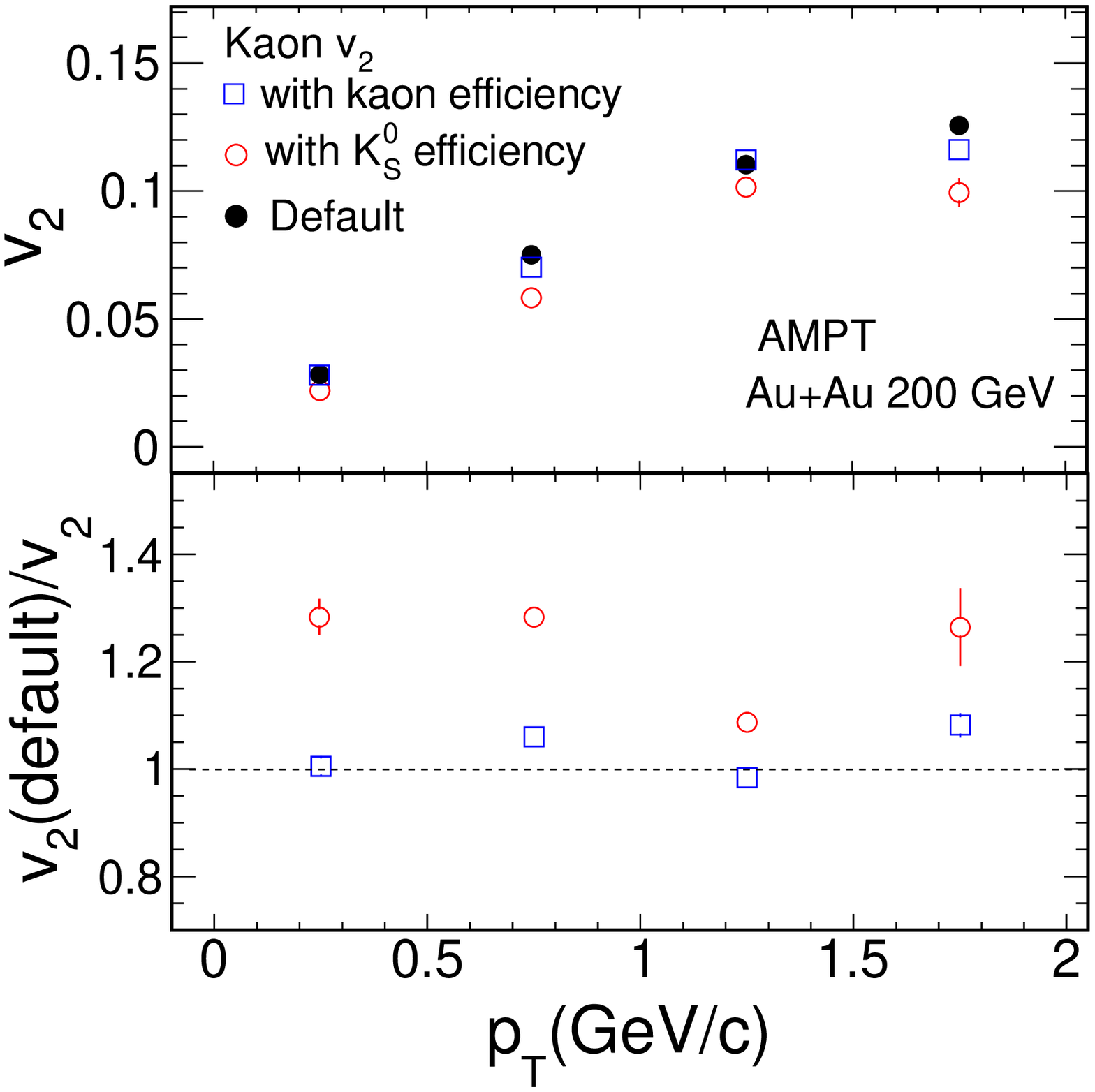}
\caption{(Color online) $v_{2}$ of kaon as function of $p_{T}$ in
  Au+Au collision for 0-80$\%$ from AMPT model  in three different
  condition: 100$\%$ kaon reconstruction efficiency (default), with
  charged kaon reconstruction efficiency and with $K^{0}_{S}$
reconstruction efficiency. }
\label{fig3}
\end{center}
\eef

\section{Effect of Centrality Determination Procedure}
\label{sec:3}
Determination of centrality selection in experiments is found to play
an important role in measurements related to particle
correlations~\cite{Adamczyk:2013dal}. Specifically when the same particles that are used for the
correlation studies also forms a subset of the particles used to obtain the
centrality selection.  Hence it is important to study the effect of 
centrality determination procedure on measurement of $v_{2}$. Further
different experiments use different methods of centrality selection.
Hence such a study is necessary to see if measured $v_{2}$ values across
different experiments can be compared. 

The simulated charged particle average $v_{2}$ ($\langle v_{2} \rangle$) values are obtained for three different ways for centrality
selection. They are: (a) centrality obtained using charged particle multiplicity
within $|\eta|$$<$ 0.5 (labeled as centrality 1),  centrality obtained using charged particle multiplicity
within $|\eta|$$>$ 0.5 and $|\eta|$$<$ 1.0 (labeled as centrality 2), and
that obtained using spectator neutrons (labeled as centrality 3). Experiments at the Relativistic Heavy-Ion Collider facility
commonly uses these methods to select on collision centrality.  Figure~\ref{fig4} show
$\langle v_{2} \rangle$ of charged particles measured at midrapidity
($|\eta|< 0.5$) as function of centrality  for the above three different
cases using AMPT model. We observed that the maximum difference in
$\langle v_{2} \rangle$  due to different
centrality selection procedure is $\sim$ 2$\%$. The agreement between
the $\langle v_{2} \rangle$ result from centrality 1 and centrality 2
shows there  is no auto-correlation effect, due to using common set of particles
for both the centrality determination and $\langle v_{2} \rangle$ calculation.
\bef
\begin{center}
\includegraphics[scale=0.45]{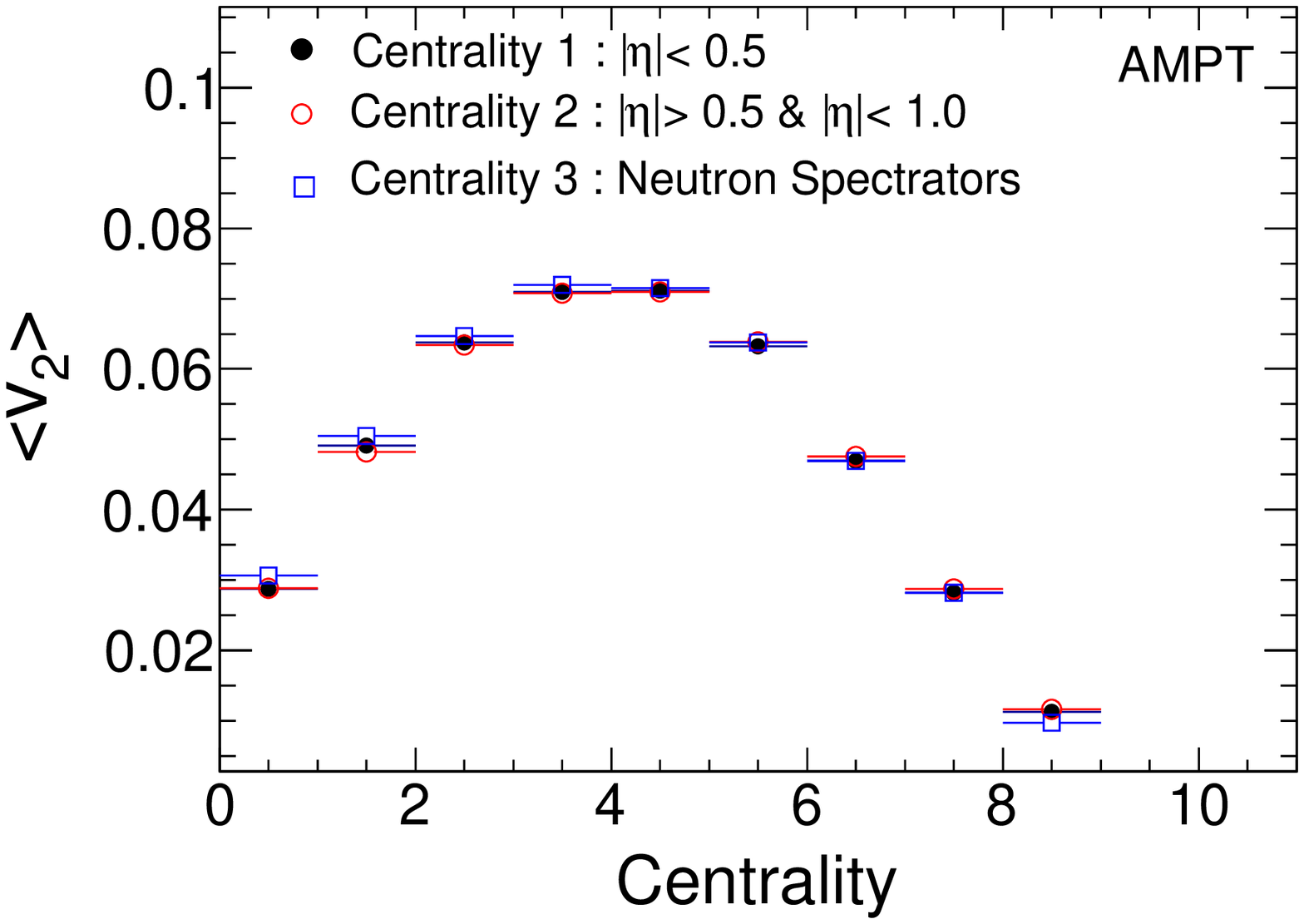}
\caption{(Color online) $\langle v_{2} \rangle$ of charged particles as
function of collision centrality at midrapidity from AMPT model. The
x-axis value of zero corresponds to most peripheral collisions (70-80\%) and 9
corresponds to most central (0-5\%) Au + Au collisions studied.}
\label{fig4}
\end{center}
\eef

\section{Event Plane Resolution Correction}
\label{sec:4}
Event plane ($\Psi$) is an estimation of true reaction plane
($\Psi_{r}$). As the estimated reaction plane
fluctuates owing to finite number of particles, one has to correct
observed $v_{2}^{obs}$ by the corresponding event plane resolution. The event plane
resolution is defined by the correlation of the event plane with the
reaction plane~\cite{Poskanzer:1998yz}:
\begin{equation}
R=\langle\cos(2(\Psi-\Psi_{r}))\rangle.
\end{equation}
The reaction plane is not measurable in experiments hence resolution
can not be calculated using above relation. To estimate the event-plane resolution we measure the correlation
between the azimuthal angles of two subset groups of tracks,
called sub-events (labeled as A and B):
\begin{equation}
R=\langle \cos(2(\Psi-\Psi_{r})) \rangle =
C. \langle \cos(2(\Psi^{A}-\Psi^{B})) \rangle.
\end{equation}
where C is factor that depends on the multiplicity of the event, $\Psi^{A,B}$
are sub event plane angles and $\langle$  $\rangle$  denote the average
over events. The resolution corrected $v_{2}$ is given as,
\begin{equation}
v_{2}
=\frac{\langle\langle\cos(2(\phi-\Psi)) \rangle\rangle}{\langle\cos(2(\Psi^{A}-\Psi^{B}))\rangle}
= \frac{v_{2}^{obs}}{R} .
\end{equation}
Most commonly used method for resolution correction for an average
$v_{2}$ over a wider centrality range (like 0-80\%) is
\begin{equation}
\langle v_{2} \rangle = \frac{\langle v_{2}^{obs}\rangle}{\langle R \rangle} .
\end{equation}
Here $\langle R \rangle$ is the mean resolution in that wide centrality bin and
can be calculated as
\begin{equation}
\langle R \rangle  = \frac{\sum N_{i} \langle R \rangle_{i}}{\sum N_{i}}.
\end{equation}
where $N_{i}$ and $\langle R \rangle_{i}$ is the multiplicity  and
resolution of the $i^{th}$ narrow centrality bin (typical centrality bin widths
of 5\% or 10\%). However as shown 
using AMPT model simulations for charged particles in Fig.~\ref{fig5}, such a procedure does
not recover back the true $v_{2}$ denoted as $v_{2}(RP)$.

Therefor another approach for event plane resolution correction for
wide centrality bin has been proposed~\cite{hmasui}. In this method
resolution correction for wide centrality bin is done by
dividing the term $\cos(2(\phi-\Psi))$ by the event plane resolution ($R$) for the corresponding centrality for each event.
\begin{equation}
\langle v_{2} \rangle = \langle\frac{v_{2}^{obs}}{R} \rangle.
\end{equation}
These two method would not yield the same value of $\langle v_{2} \rangle$ because
\begin{equation}
\frac{\langle v_{2}^{obs}\rangle}{\langle R \rangle} \neq
\langle\frac{v_{2}^{obs}}{R} \rangle.
\end{equation}
\bef
\begin{center}
\includegraphics[scale=0.45]{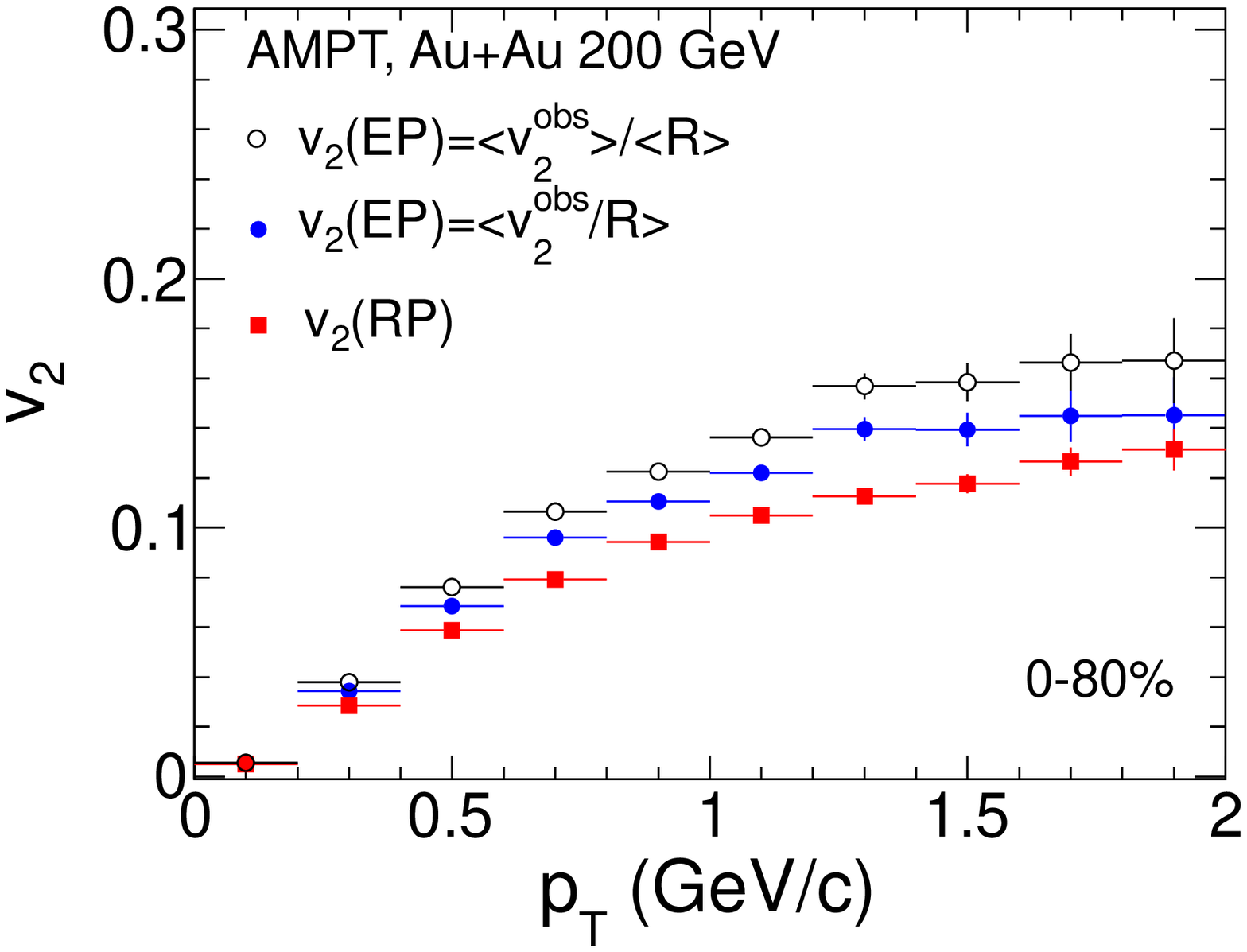}
\caption{(Color online) The $v_{2}$ of charged particles as
function of $p_{T}$ for 0-80$\%$ centrality  in Au + Au
collisions from AMPT model. The results for two different methods of event plane resolution
correction are compared to the true $v_{2}$ values obtained using the
known reaction plane angle in the model.}
\label{fig5}
\end{center}
\eef
Figure~\ref{fig5} shows charged particles  $v_{2}$ as
function of $p_{T}$ for 0-80$\%$ centrality bin in Au + Au
collisions. The red marker corresponds to $v_{2}$ measured with
respect to true reaction plane. Open black and solid blue circle
corresponds to $v_{2}$ measured with
respect to event plane and resolution correction done using method
described in equation 5 and 7, respectively. The  $v_{2}$ measured with
respect to true reaction plane is the actual $v_{2}$ in the AMPT
model. The results in Fig.~\ref{fig5} shows that event-by-event resolution correction
method using the equation 7 gives $v_{2}$ values closer to the
true $v_{2}$.

\section{Event Bias Correction}
\label{sec:5}
\begin{figure}[!ht]
\begin{center}
\includegraphics[scale=0.33]{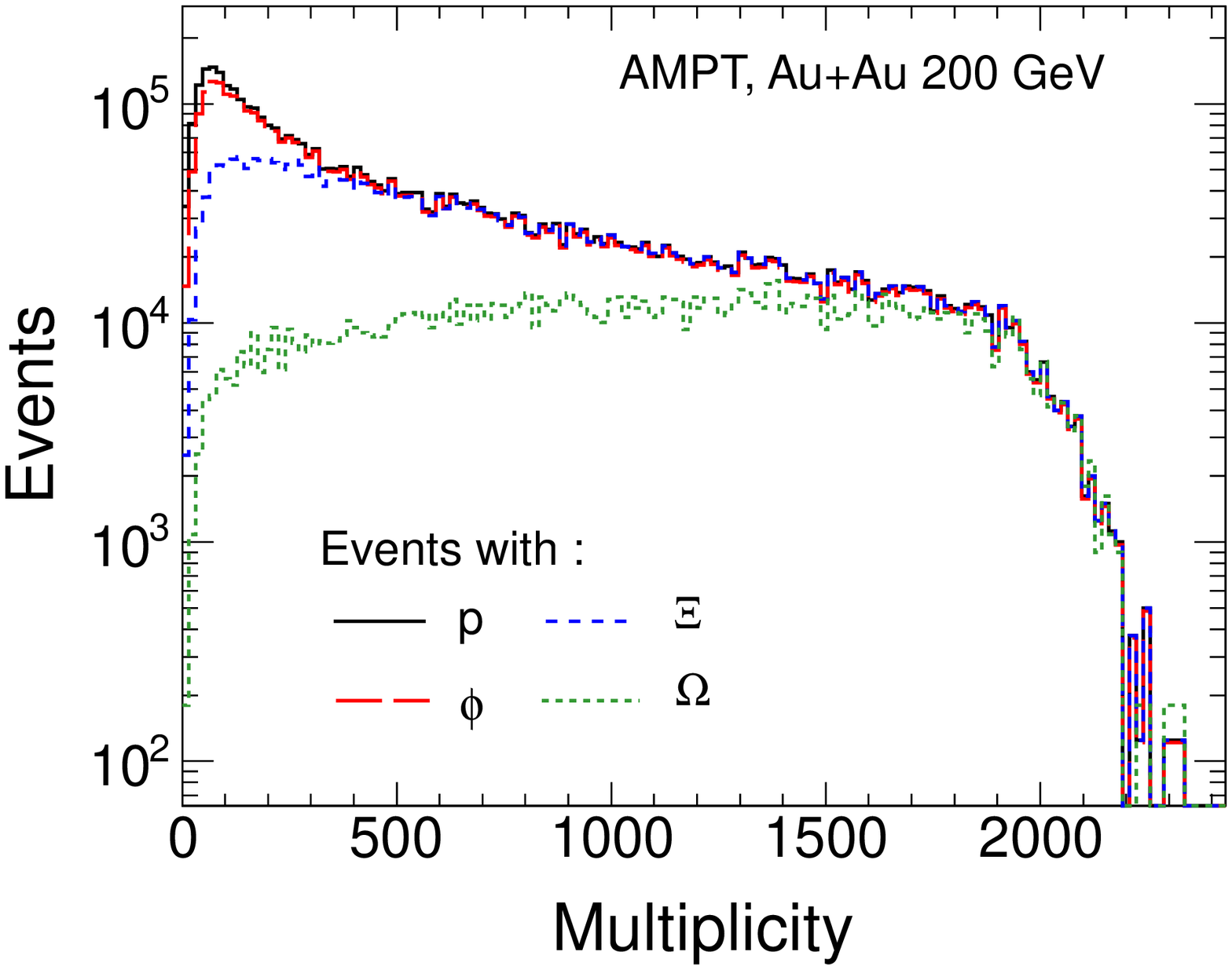}
\includegraphics[scale=0.34]{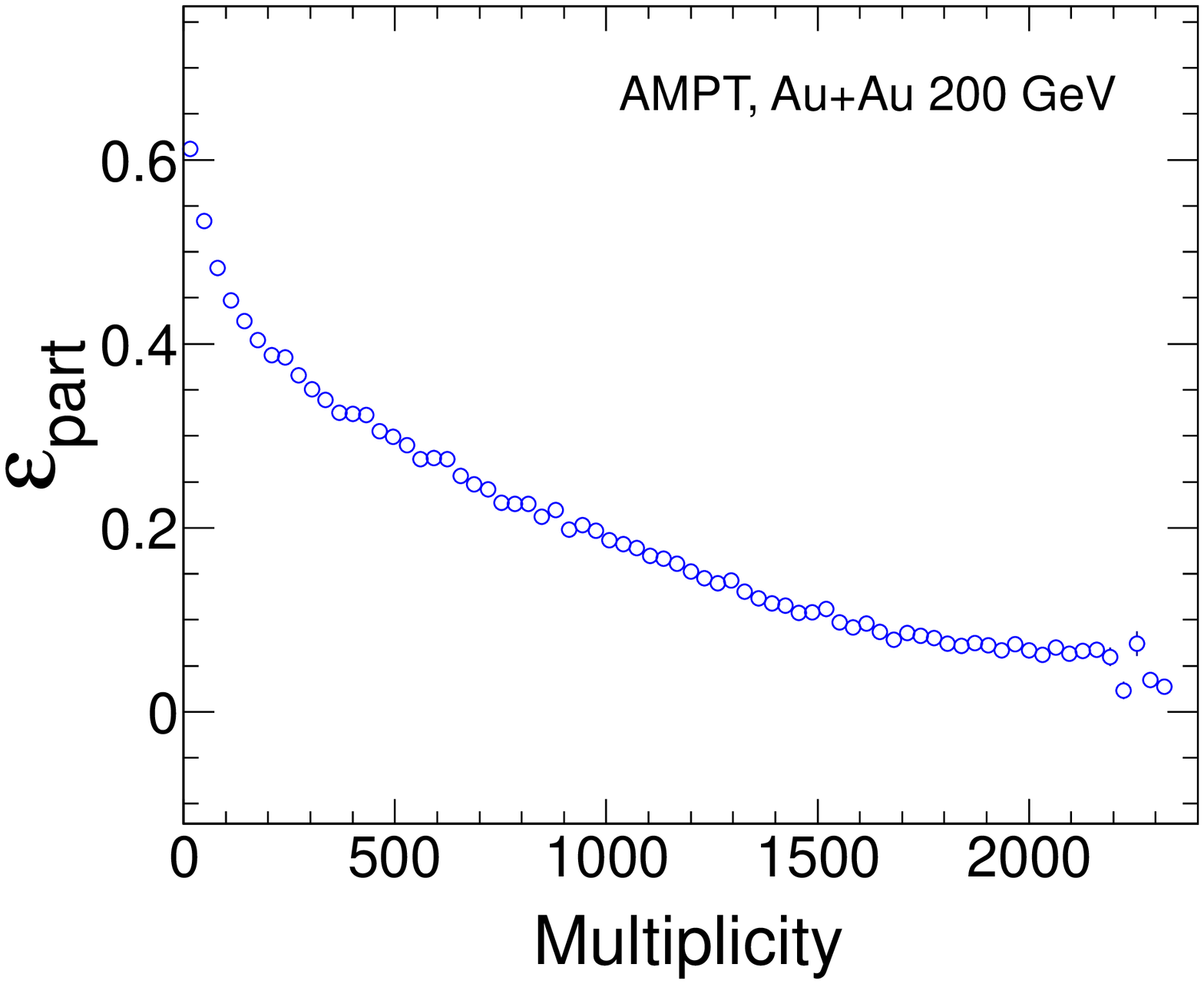}
\end{center}
\caption{(Color online) Top panel: Number of events as a function of
   particle multiplicity in Au+Au collisions at
  $\sqrt{s_{NN}}$ = 200 GeV from AMPT model. Bottom panel: Participant eccentricity
   as function of particle multiplicity in  Au+Au collisions at
  $\sqrt{s_{NN}}$ = 200 GeV from AMPT model.}
\label{msh_refmult}
\end{figure} 
In the particular case of comparing the $v_{2}$ values for heavier 
hadrons such as $\Omega$ to those copiously produced such as pions
or protons in minimum bias collisions there is an inherent bias towards the
event class. This is illustrated in the top panel of
Fig.~\ref{msh_refmult}. It shows the number of events as function of
particle multiplicity in Au+Au collisions at  $\sqrt{s_{NN}}$ = 200 GeV from
AMPT Model. The black, red, blue and green  histogram corresponds to
the events which contain at least one proton, $\phi$, $\Xi$ and
$\Omega$, respectively. The minimum bias multiplicity distribution 
with heavier hadrons like $\Xi$ and $\Omega$ are very different from
those for protons. The participant eccentricity ($\varepsilon_{part}$) obtained
using the initial position of the participating
nucleons~\cite{Alver:2010gr,Haque:2011ti} is
correlated with the particle multiplicity as seen from
Fig.~\ref{msh_refmult} bottom panel. Hence the average $\varepsilon_{part}$ of
events containing multi-strange hadron like $\Omega$ in 0-80\% wide
centrality would be smaller than the average $\varepsilon_{part}$ determined for
events containing protons. Since $v_{2}$ is driven by the anisotropy of the initial
spatial geometry, therefore the event bias is naturally introduced when
comparison is made between $v_{2}$ measured in a wide centrality bin
especially for the rarely produced particles like $\Omega$ to that for
protons. This event bias effect needs to be corrected before
comparisons of minimum bias $v_{2}$ values for different types of
hadrons. This bias could be corrected by normalising the measured 
$v_{2}$ by the ratio of standard average $\varepsilon_{part}$  (for charged
particle) to the average $\varepsilon_{part}$ of the events which contains the particle
of interest weighted by the corresponding yield. The correction factor
in the present calculations using AMPT data for $\Omega$  is 1.15
where as for $\Xi$ and $\phi$ it is 1.023 and 1.010, respectively.
The lighter hadrons do not have a large event bias correction, due to
their copious production in nuclear collisions at RHIC energy. 

\begin{figure}[!ht]
\begin{center}
\includegraphics[scale=0.33]{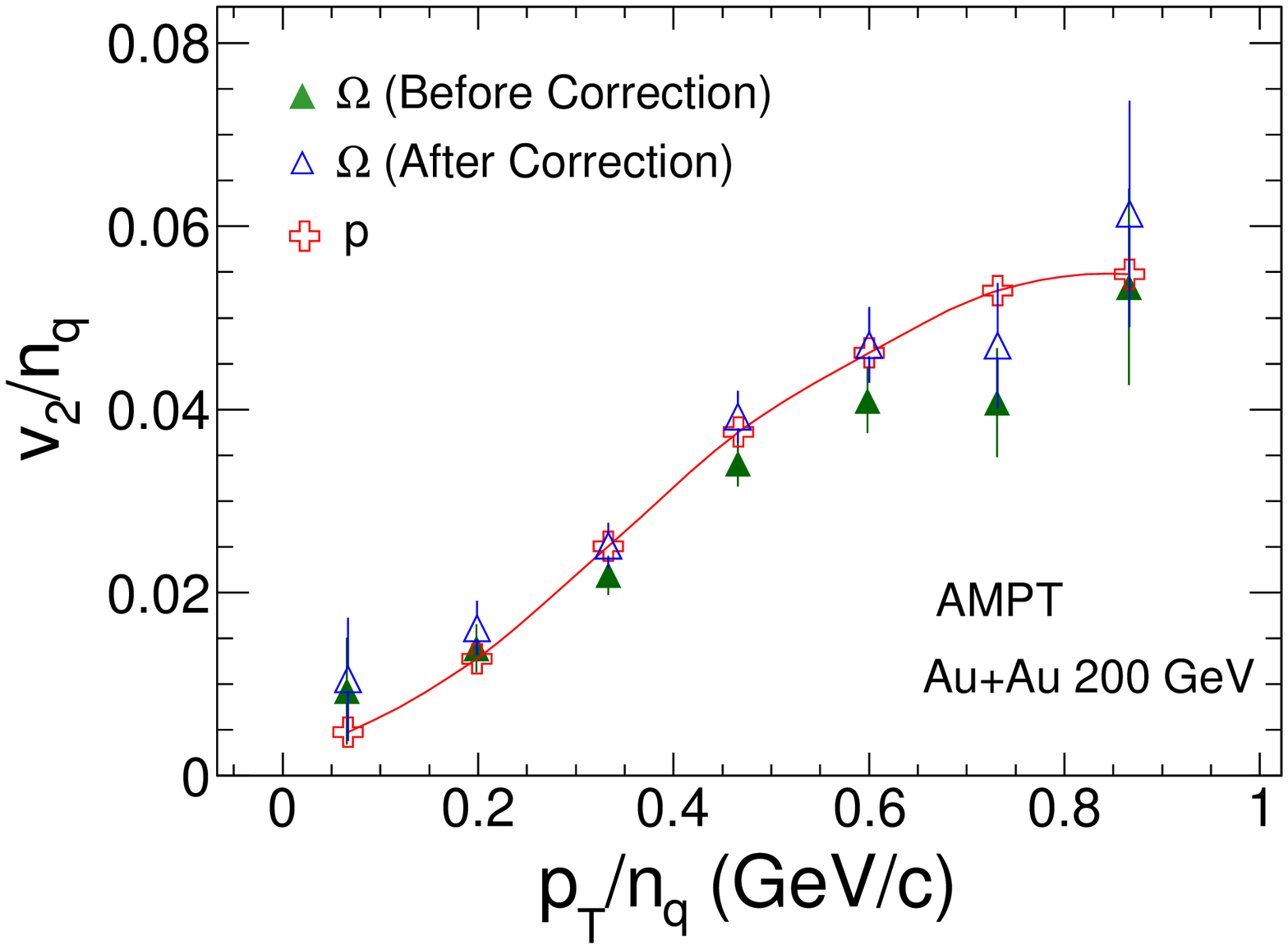}
\end{center}
\caption{(Color online) $v_{2}/n_{q}$ as a function of $p_{T}/n_{q}$
  in 0-80\% minimum bias Au+Au collisions ($|\eta| < 0.5 $) at
  $\sqrt{s_{NN}}$ = 200 GeV from AMPT  model. $n_{q}$ is the number
  of constituent quarks, equal to 3 here for the baryons.}
\label{ncq_plot}
\end{figure} 
The physical consequence of such an event bias is shown in
Fig.~\ref{ncq_plot}. The number of constituent quark scaling between
$\Omega$ and proton which is naturally expected in AMPT  model 
holds better at the intermediate $p_{T}$ only after the event bias
correction as described above.

\section{Resonance Decay Effect}
\label{sec:6}
In the heavy-ion collision a large fraction of stable hadrons are from resonance decays. To study the
effect of resonance decays on the elliptic flow of stable hadrons, we
have used  the UrQMD model. 

\begin{figure}[h]
\includegraphics[scale=0.42]{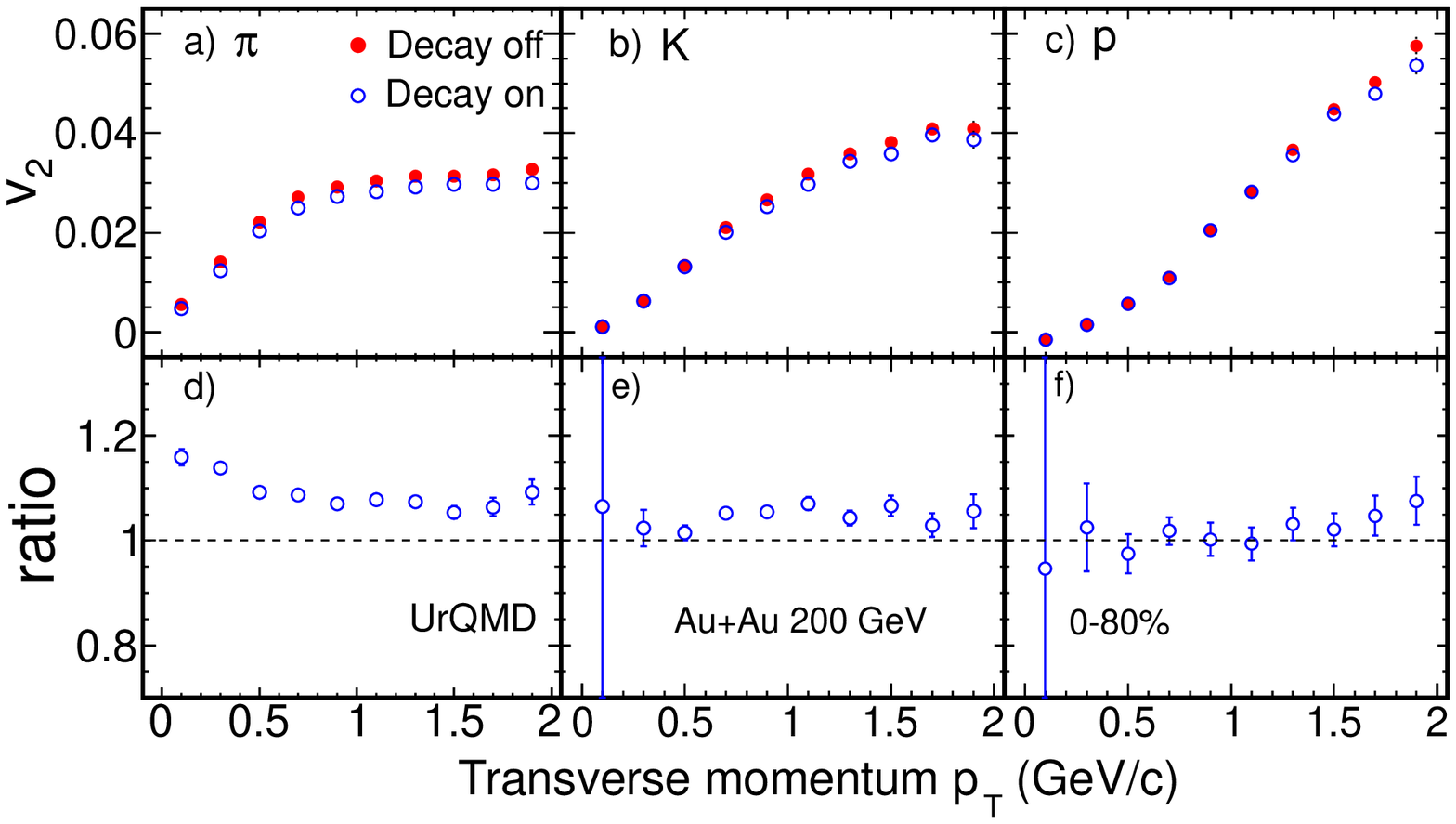}
\caption{(Color online) The $v_{2}$ of $\pi$, $K$ and $p$ as
function of $p_{T}$ at 0-80$\%$ centrality with decay off and decay on
condition from minimum bias Au+Au
collision at $\sqrt{s_{NN}}$ = 200 GeV from UrQMD model.}
\label{fig6}
\end{figure}

Specifically we study the effect of resonances, such as $\rho$,
$\Lambda$, $\phi$, $\eta$, $\Omega$, $\Sigma$ and $\Delta$ on measured
$v_{2}$ of inclusive pion, kaon and proton. In the
UrQMD model there is a option for switching off and on the decay of each
resonance. Figure~\ref{fig6} shows $v_{2}$ of $\pi$, $K$ and $p$ as
function of $p_{T}$ with decay off and decay on condition in Au+Au
collision from UrQMD model. The ratios shown in the lower panels
of Fig~\ref{fig6}, shows that  there is change of 10$\%$ to
15$\%$ in the $v_{2}$ values of pion, less than 5$\%$ for the kaon and
the $v_{2}$ values for the proton is almost unaffected. There is a decrease in
$v_{2}$ values due to the decay of resonances. However, one could expect a
higher value of $v_{2}$ from decay of resonances. The decay particle
at given transverse momentum arises mostly from a resonance at higher
momentum. The $v_{2}$ value in general increases with $p_{T}$.
However, it seems the decay process being isotropic in the
rest frame of the resonance, reduces the $v_{2}$~\cite{rho_decay}. To understand
the results better,  we have further studied the effect from the decay of
$\rho$ $\longrightarrow$ $\pi^{+}$ + $\pi^{-}$ on $v_{2}$ of
pions. This decay process is isotropic in the
rest frame of the resonance and hence one can expect reduction in the
momentum anisotropy of the daughter pions. 

\begin{figure}[!h]
\includegraphics[scale=0.4]{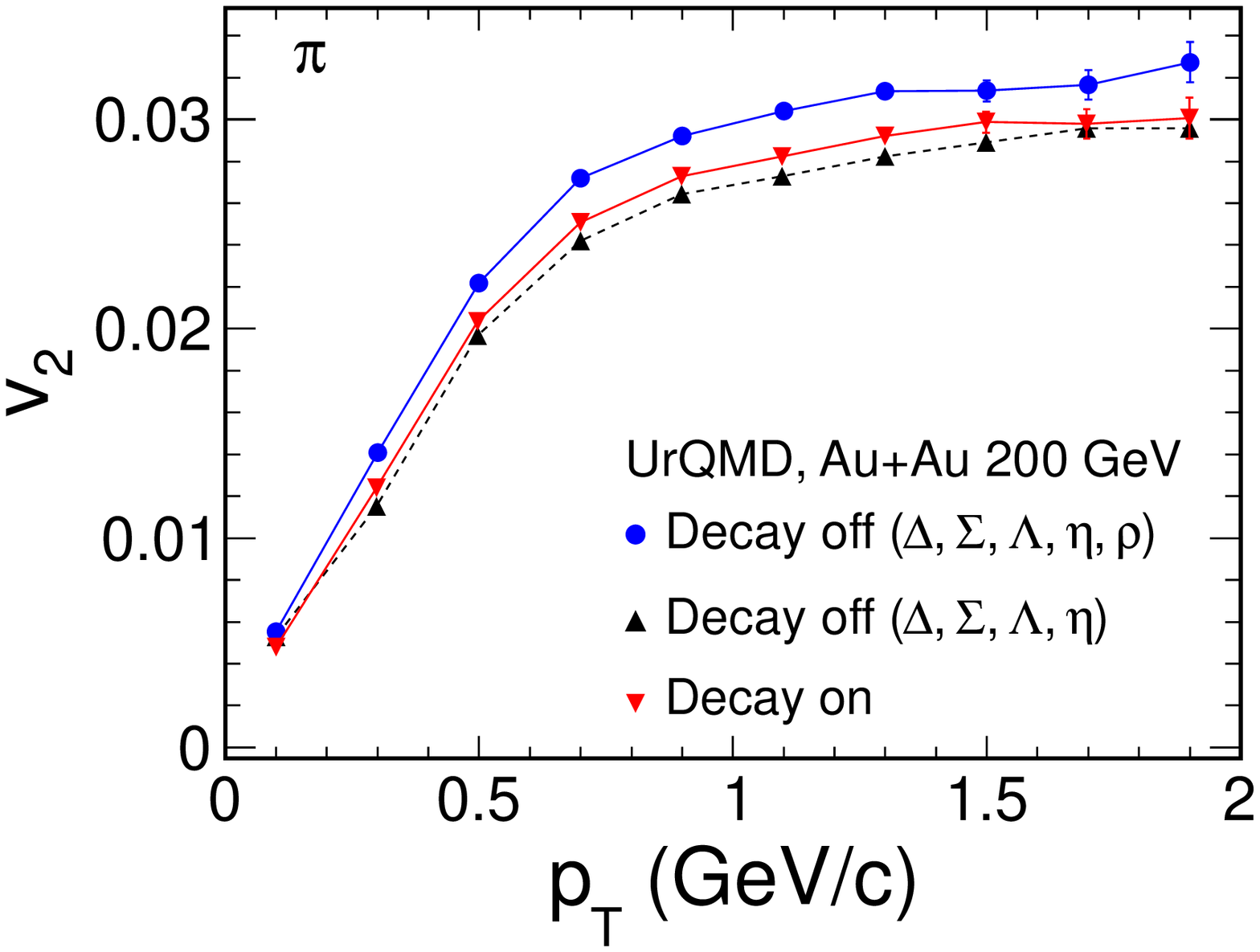}
\caption{(Color online) $v_{2}$ of pion as a function of $p_{T}$ in
Au+Au collisions at $\sqrt{s_{NN}}$ = 200 GeV for  0-80$\%$ centrality
from UrQMD model. The different results shown corresponds to different
contributions from resonance decay.}
\label{fig7}
\end{figure}

Figure~\ref{fig7} shows the $v_{2}$ of pion as a function of $p_{T}$ in
Au+Au collisions at $\sqrt{s_{NN}}$ = 200 GeV from UrQMD model for three different cases: (a)
All resonances are decayed (shown by red inverted triangles), (b)  $\rho$, $\Lambda$,
$\eta$, $\Sigma$ and $\Delta$ are not decayed (shown by solid blue circle)
and,  (c)  $\Lambda$,
$\eta$,  $\Sigma$ and $\Delta$ are not decayed (shown by black triangle). There is
decrease in the $v_{2}$ values of pion due to the decay of $\rho$ $\longrightarrow$
$\pi^{+}$ + $\pi^{-}$ as expected from the decay kinematics
(comparison between cases (b) and (c)). However, the pion $v_{2}$ values
increases due to decay of $\Lambda$,
$\eta$,  $\Sigma$ and $\Delta$ (comparing between cases (a) and (c)).
Similarly we have observed that the decreases in kaons $v_{2}$ is due to $\phi$ $\longrightarrow$
$K^{+}$ + $K^{-}$ decay (not shown here). There is an overall decrease
in the measured value of pion and kaon $v_{2}$ due to the decay of the
resonances.

\section{Elliptic flow from Scalar Product method}
\label{sec:7}
In a scalar product method~\cite{sp} of determining $v_{2}$ each event is partitioned
into two sub events, labeled here as A and B. If $Q_{n}^{A}$ and $Q_{n}^{B}$ are the flow vectors of sub events A
and B for the $n^{th}$ harmonic then the correlation between the two
sub events is given as
\begin{equation}
\langle Q_{n}^{A} Q_{n}^{B*}\rangle = \langle v_{n}^{2} M^{A}M^{B} \rangle,
\end{equation}
where $M^{A}$ and $M^{B}$ are the multiplicities for sub events A and
B, respectively. Elliptic flow parameter in this method is given as
\begin{equation}
v_{2}(SP) = \frac{\langle Q_{2}u_{2*} \rangle}{\sqrt{ \langle Q_{2}^{A} Q_{2}^{B*}\rangle} }.
\end{equation}
Here $Q_{2} = \sum u_{2}^{i}$
and $u_{2}^{i}$ is a unit vector associated with the $i^{th}$
particle. The scalar-product method
\begin{figure}[h]
\includegraphics[scale=0.38]{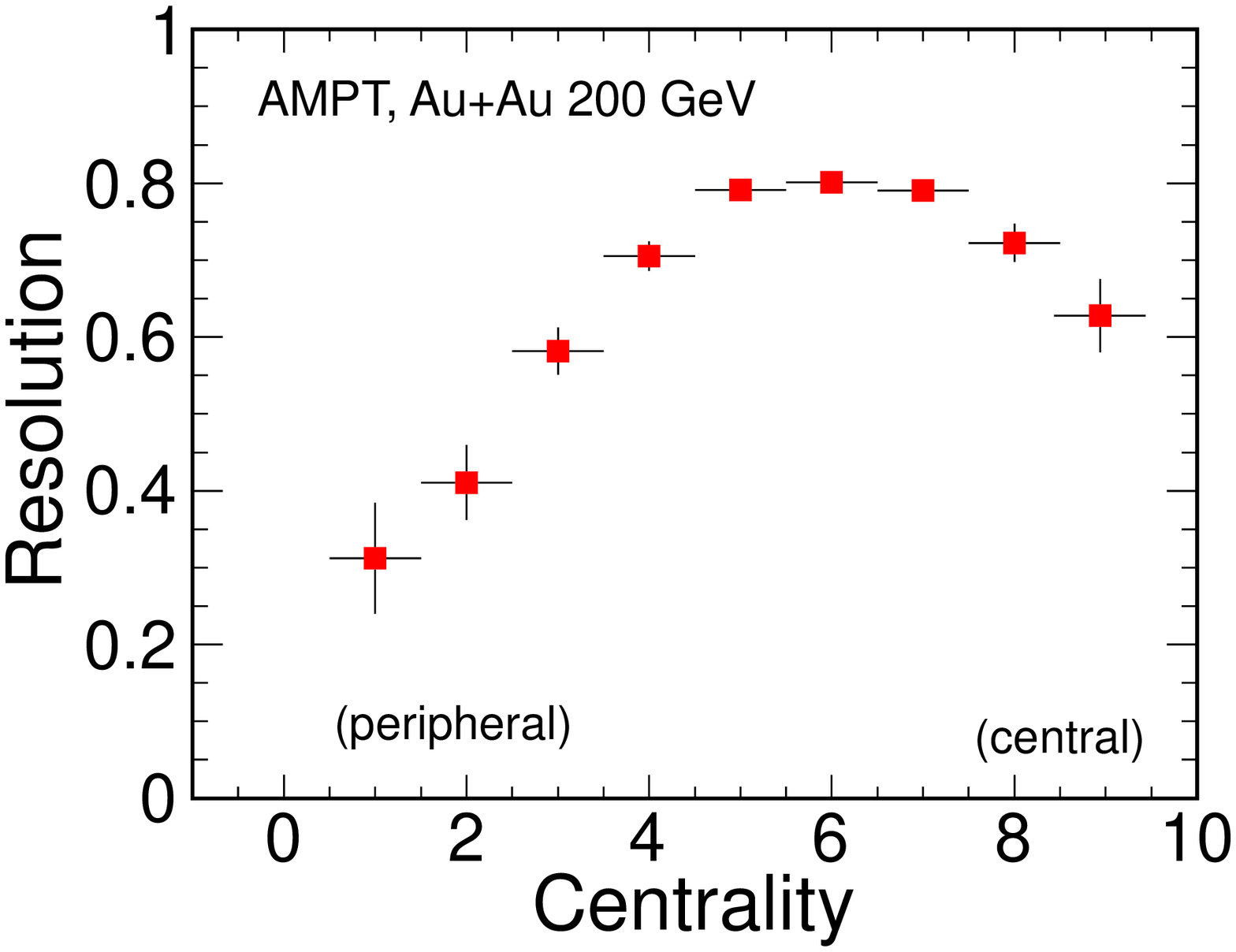}
\caption{(Color online) Second order event plane resolution as a function of centrality in
Au+Au collisions at 200 GeV from AMPT model.}
\label{fig8}
\end{figure}
\begin{figure}
\includegraphics[scale=0.38]{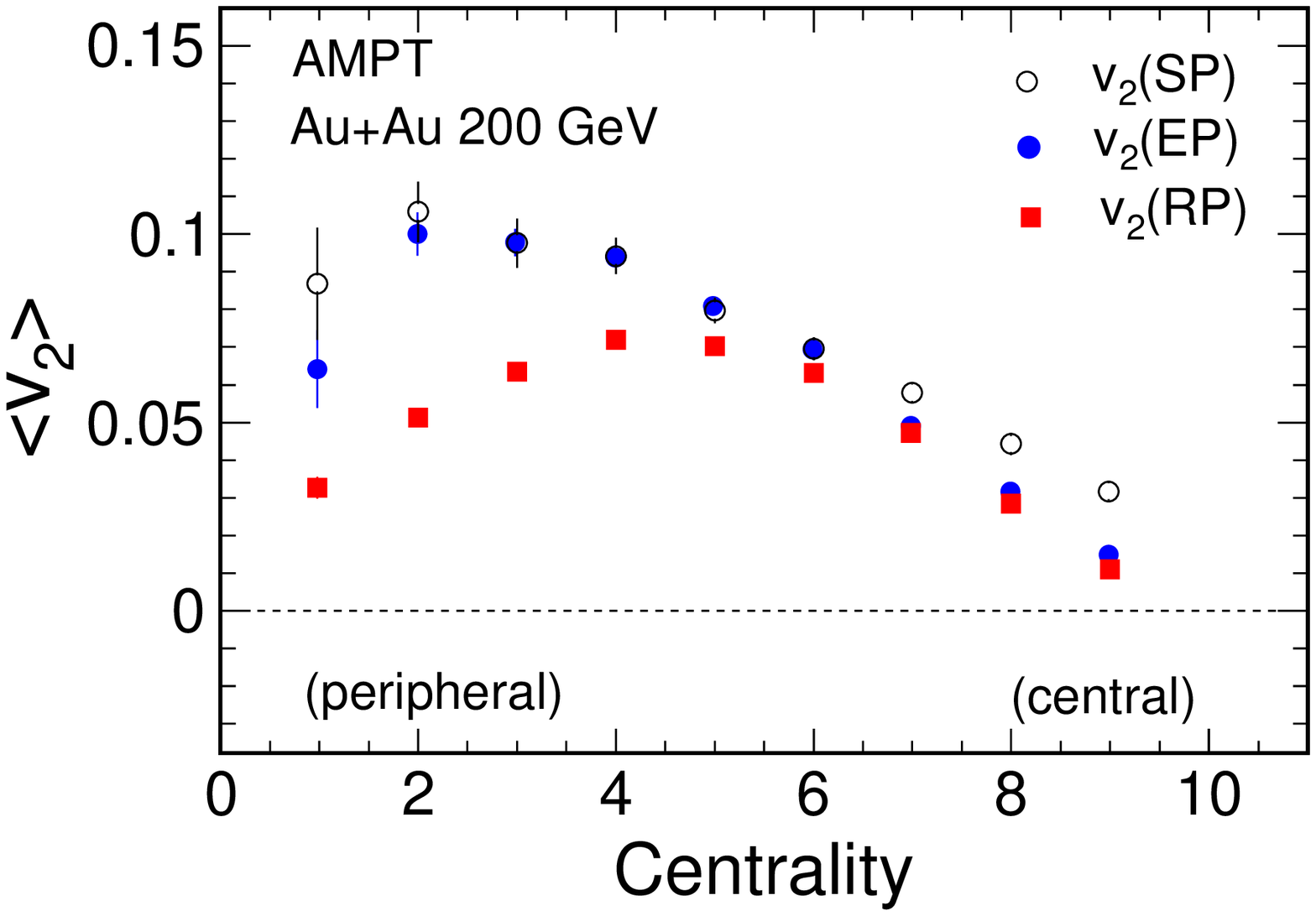}
\caption{(Color online) The elliptic flow of charged particle obtained
  using different methods as a function of centrality in
Au+Au collisions at 200 GeV  from AMPT model.}
\label{fig9}
\end{figure}
always yields the root-mean-square $v_{2}$, regardless of the details
of the analysis~\cite{Luzum:2012da}.
\begin{equation}
v_{2}(SP) = \frac{\langle Q_{2}u_{2*} \rangle}{\sqrt{ \langle Q_{2}^{A} Q_{2}^{B*}\rangle} }  = \sqrt{\langle v_{2}^{2} \rangle }.
\end{equation}
But this is not true for $v_{2}(EP)$ measured by  conventional event
plane method. Recently it has been argued~\cite{Luzum:2012da} that in the limit
of perfect resolution (i.e. $R$ $\longrightarrow$ 1 ) 
\begin{equation}
v_{2}(EP) \longrightarrow \langle v_{2} \rangle,
\end{equation}
and in the limit of low resolution 
\begin{equation}
v_{2}(EP) \longrightarrow \sqrt{\langle v_{2}^{2} \rangle }.
\end{equation}
We have investigate this using AMPT model where the actual $\langle
v_{2} \rangle$ in known ($v_{2}(RP)$). The event plane resolution from AMPT model is
shown in Fig.~\ref{fig8} for nine centrality bins, corresponding to
0-5\%, 5-10\%, 10-20\%, 20-30\%, 30-40\%, 40-50\%, 50-60\%, 60-70\%
and 70-80\% for the cross section. Resolution is poor for the peripheral centrality and
maximum at mid-central and then slightly
decreases for the most central collisions. This is due to the
interplay of multiplicity and $v_{2}$ for different centrality bins. Figure ~\ref{fig9} shows charged
particle $v_{2}$ as a function of centrality in Au+Au collisions at
200 GeV  from AMPT model using scalar product, event plane and
reaction plane methods. The most central collisions studied corresponds to a value
of 9 in the $x$-axis and most peripheral collisions corresponds to a value
of 0 in the $x$-axis. For the peripheral collision
where resolution is poor, $v_{2}(EP)$ and $v_{2}(SP)$ are very close
to each other  that means $v_{2}(EP)$ is equivalent to $\sqrt{\langle v_{2}^{2} \rangle}$. However for central to mid-central collisions where
resolution is high,  $v_{2}(EP)$ is closer to $v_{2}(RP)$ or $\langle v_{2} \rangle$. The results are consistent with with equations 12-13
and as proposed in Ref.~\cite{Luzum:2012da}.

\section{Summary}
\label{sec:8}
We have presented a transport model based study of various effects on
experimentally measured $v_{2}$. The results are presented using Au+Au
collisions at midrapidity at $\sqrt{s_{NN}}$ = 200 GeV from AMPT and
UrQMD model. We find that finite particle counting efficiency of the
detectors used in real experiments affects the measured $v_{2}$
values. Specifically due to the difference in the reconstruction
efficiencies of charged kaon and neutral kaons, the measured $v_{2}$
values could differ by 10-30\% as a function of $p_{T}$. Experiments
needs to account for this inefficiencies while obtaining $v_{2}$ using
event plane method, before comparing the results
for measured hadrons with very different reconstruction efficiencies. 
We observe that the measured $v_{2}$ values remain insensitive to the
method of centrality determination used in the experiments. However
the procedure to correct for event plane resolution in wide centrality
bin results affects the extracted $v_{2}$ values. Event-by-event
resolution correction seems to give $v_{2}$ values that are closer to
the true  $v_{2}$  values. The over all effect of resonance decay is
to reduce the $v_{2}$ values relative to the true $v_{2}$. This is
dominated due to kinematic effect of the decay process being isotropic
in the rest frame of the resonance and such resonances contributing
more in terms of the yields of the measured hadrons. The minimum bias event class in
terms of average initial $\varepsilon_{part}$ value for events having rare heavier
particle like $\Omega$ is different from those having protons. In
order to appropriately compare the minimum bias $v_{2}$ values of
various hadrons, we propose an event bias correction procedure. We
also demonstrate that this procedure seems to work by showing that
number of constituent quark scaling for $\Omega$ and protons $v_{2}$.
Finally we have
demonstrated through the model study that in the limit of high
resolution for event plane determination the $v_{2}(EP)$ $\longrightarrow$
$\langle v_{2} \rangle$ and in the limit of small event plane resolution the
$v_{2}(EP)$ $\longrightarrow$ $\sqrt{\langle v_{2}^{2} \rangle}$.

\begin{acknowledgements}
Financial assistance from the SwarnaJayanti Fellowship of the Department of Science and Technology, Government 
of India is gratefully acknowledged. We thank Dr. H. Masui, Dr. S. Shi
and Dr. N. Xu for useful discussions on event bias correction method.
\end{acknowledgements}



\end{document}